\documentclass[12pt]{JHEP3} 

\usepackage{mathrsfs}
\usepackage{amsmath,amssymb}

\renewcommand{\o}{{\omega}}

\def \x {\times}
\def \a {\alpha}
\def \pa{\partial}
\def\w{\omega}
\def\b{\beta}
\def\l{\lambda}
\def\eps{\epsilon}

\def\s{\sigma}

\newcommand{\be}{\begin{equation}}
\newcommand{\ee}{\end{equation}}
\newcommand{\bea}{\begin{eqnarray}}
\newcommand{\eea}{\end{eqnarray}}

\newcommand{\alg}[1]{\mathfrak{#1}}

\def\L{\mathscr{L}}

\newcommand{\su}{\alg{su}}

\newcommand{\GA}{G_{tt}}
\newcommand{\GS}{G_{\phi\phi}}
\newcommand{\AdS}{{\rm  AdS}_5\times {\rm S}^5}
\newcommand{\Z}{\mathcal Z}
\newcommand{\Y}{\mathcal Y}
\newcommand{\V}{\mathcal V}
\def\S{\rm S}

\title{
Integrable Hamiltonian \\
for Classical Strings on ${\rm\bf  AdS}_5\times {\rm\bf S}^5$
}

\author{Gleb Arutyunov$^{1,\dagger}$ and Sergey Frolov$^{2,\dagger}$ \\
\vskip 0.5cm
$^{1}$Max-Planck-Institut f\"ur Gravitationsphysik,
Albert-Einstein-Institut\\
Am M\"uhlenberg 1, D-14476 Potsdam, Germany\\
\vskip 0.1cm
$^{2}$Department of Applied Mathematics, \\
SUNY Institute of Technology,\\
P.O. Box 3050, Utica, NY 13504-3050\\

\vskip 0.1cm
Email: \email{agleb@aei.mpg.de; frolovs@sunyit.edu}   }

\preprint{\hepth{0411089}\\
AEI-2004-105}

\abstract{We find the Hamiltonian for physical excitations of the
classical bosonic string propagating in the ${\rm AdS}_5\times
{\rm S}^5$ space-time. The Hamiltonian is obtained in a so-called
uniform gauge which is related to the static gauge by a 2d
duality transformation. The Hamiltonian is of the Nambu type and
depends on two parameters: a single ${\rm S}^5$ angular momentum
$J$ and the string tension $\lambda$. In the general case both
parameters can be finite. The space of string states consists
of short and long strings. In the sector of short strings the
large $J$ expansion with $\lambda'=\frac{\lambda}{J^2}$ fixed
recovers the plane-wave Hamiltonian and higher-order corrections
recently studied in the literature. In the strong coupling limit
$\l\to \infty$, $J$ fixed, the energy of short strings scales as
$\sqrt[4]{\l}$ while the energy of long strings scales as
$\sqrt{\l}$. We further show that the gauge-fixed Hamiltonian is
integrable by constructing the corresponding Lax representation.
We discuss some general properties of the monodromy matrix, and
verify that the asymptotic behavior of the quasi-momentum
perfectly agrees with the one obtained earlier for some specific
cases.

\vskip 0.5cm $^{\dagger}$ Also at Steklov Mathematical Institute,
Moscow. }


\begin{document}

\section{Introduction}

Our understanding of the gauge/string duality \cite{M} has
recently improved due to new ideas and techniques on both sides of
the AdS/CFT correspondence. Even though the duality is of
strong/weak coupling type, it was proposed  by Berenstein,
Maldacena and Nastase (BMN) \cite{BMN} that energies of certain
string states can be matched with perturbative scaling dimensions
of dual SYM operators. The BMN results were re-interpreted in
\cite{GKP2} as the semi-classical quantization near a point-like
string carrying large momentum $J$ along the central circle of
${\rm S}^5$. The BMN proposal was then generalized in \cite{FT} where it
was found by using the semi-classical approach that there exists a
large sector of highly energetic string states on $\AdS$ which permits
a direct comparison with perturbative gauge
theory. Any quantum ``heavy'' state can be well approximated by a
classical string. However, in most cases the string energy turns
out to be non-analytic in the 't Hooft coupling constant $\lambda$
\cite{GKP2}, that precludes a direct comparison with perturbative
gauge theory.

Quite remarkably, as was shown in \cite{FT}, energies of classical
multi-spin strings rapidly rotating in $\S^5$ admit an expansion
in integer powers of the effective coupling constant
$\lambda/L^2$, where $L$ is the large, total spin on $\S^5$. Even
though the coupling $\lambda$ is large in the semi-classical
approach, this expansion is of the same form as in perturbative
gauge theory, and, therefore, one may compare the energies of
spinning strings with perturbative scaling dimensions of gauge
theory operators.

The multi-spin string solutions are completely determined by the
bosonic part of the Green-Schwarz superstring action \cite{MT1}.
In the conformal gauge the bosonic string is described by the
sigma model with the $\AdS$ target space, which is known to be
exactly solvable. A finite-dimensional reduction of the classical
string sigma model to an integrable system of the Neumann type
\cite{AFRT} describes folded and circular rigid strings. More
general string solutions are described by integral equations of the Bethe type.
For strings moving in ${\mathbb R}\times
\S^3$, ${\rm AdS}_3\times \S^1$ and ${\mathbb R}\times \S^5$ they were
derived in \cite{KMMZ}.

A related development in gauge theory was triggered by an
important observation \cite{MZ} that in the ${\rm SO}(6)$ subsector
planar superconformal ${\cal N}=4$ SYM is an integrable system in
the one-loop approximation. This was generalized to the complete
dilatation operator in \cite{BS} and, the integrability, very
likely, holds at higher loops as well \cite{BKS}. Integrable
structures of QCD were previously observed in \cite{BDM}.
Integrability allows one to formulate a system of Bethe equations
which is then used to find anomalous dimensions of conformal
operators. For a closed $\su(2)$
subsector the one-loop Bethe ansatz of \cite{MZ} is
extended up to three loops \cite{SS} owing to the fact that
the corresponding three-loop dilatation operator \cite{BKS} can be
embedded into the Inozemtsev long-range spin chain \cite{Inoz}.
Recently the all-loop asymptotic Bethe ansatz for the dilatation operator
acting in the $\su(2)$ subsector was proposed in \cite{BDS}.

The spin chain Bethe equations were used in \cite{BFMSTZ,SS} to
demonstrate one- and two-loop agreement between gauge and string
theory predictions in the cases of folded and circular rigid
strings. Moreover, as was shown in \cite{AS}, the eigenvalues
of higher local commuting charges also agree up to two-loop order, indicating
a close relation between integrable structures of gauge and string theories.
Furthermore, up to the second order of perturbation theory,
the string Bethe equations
\cite{KMMZ} coincide with the spin chain Bethe equations \cite{SS}
thus leading to a
proof of two-loop agreement of string and gauge theory results in
the $\su(2)$ subsector.  The one- and two-loop agreement in
various subsectors of the gauge theory was also demonstrated in
\cite{Upp,Kruc,KrTs1, KrTs2, Kop}. The other relevant aspects of
the gauge/string duality have been investigated in \cite{Rus,nloc}.

At two leading orders of perturbation theory
the matching between gauge and string theory quantities
was also observed for $1/L$ corrections to the BMN limit
\cite{PR,Callan}. It was noticed, however, that it breaks down
at three-loop order \cite{Callan}. The same pattern, {\it i.e.} the one- and two-loop
agreement and disagreement starting at three loops,
was also found for spinning strings \cite{SS,AS}.
Moreover, results of
\cite{LZ,FPT} indicate that $1/L$ corrections for spinning
strings would disagree already at the one-loop level. As was stressed in
\cite{SS,BDS}, the origin of all these disagreements could be due to neglecting on
the gauge theory side
the so-called wrapping interactions, and the
agreement might be restored after they would have been properly incorporated.
Let us also note that disagreement between gauge and string theories becomes manifest
if one compares the thermodynamic limit of the all-loop
asymptotic  Bethe ansatz \cite{BDS} describing (infinitely) long
operators, and the string Bethe equations \cite{KMMZ} describing
classical spinning strings.

Assuming the validity of the AdS/CFT correspondence one should
expect existence of a Bethe ansatz for quantum strings which would
serve as a discretization of the integral (continuous) Bethe
equations for classical strings and, from the gauge theory
perspective, include terms responsible for wrapping interactions.
An interesting discretization of the string Bethe equations was
recently proposed in \cite{AFS}. The Bethe ansatz for quantum
strings \cite{AFS} reproduces the near BMN spectrum of
\cite{Callan}, the famous $\sqrt[4]{\lambda}$ behavior at strong
coupling \cite{GKP1}, and has a spin chain description at weak
coupling
\cite{Beissp}. The general multi-impurity spectrum (in the
$\su(2)$ subsector) predicted in \cite{AFS} has been recently
reproduced from the quantized string theory in the near plane-wave
background \cite{McL}. Existence of a Bethe ansatz with such
remarkable properties provides a strong evidence in favor of
integrability of quantum strings. An interesting recent discussion
of the quantum integrability for strings in the near plane-wave
background ($1/J$ order) can be found in \cite{Swan2}.

A necessary (but not always sufficient) condition for classical
integrability of a solvable model is the existence of a
Lax pair. The Lax representation implies the existence of an
infinite number of local
conserved charges that
may allow one to solve the system exactly. A Lax pair for the
classical superstring theory on $\AdS$ was found in \cite{BPR}
(see also \cite{Hatsuda}). To analyze quantum integrability of the
superstring theory one would need to develop the Hamiltonian
formalism. The Poisson bracket of the $\L$-operator entering the
Lax pair determines a classical $r$-matrix which is further used to
find the Poisson algebra of the corresponding monodromy operator,
and to quantize the
model. The knowledge of the Hamiltonian structure is also
crucial to exhibit commutativity of local
integrals of motion.
Unfortunately, the Lax pair of \cite{BPR} cannot be
immediately used to find an $r$-matrix structure and, therefore, to quantize the model
because the local $\kappa$-symmetry has not been fixed, and,
as the consequence, the Poisson structure of the $\L$-operator remains to be
undefined. Also, to find the Lax pair, the two-dimensional metric on the string
world-sheet has been fixed in a way equivalent
to fixing the conformal gauge. It is well-known that in the
conformal gauge fixing  $\kappa$-symmetry
leads to a very complicated Poisson structure for fermions which
can be hardly used to quantize superstring even in flat space. Of course,
the standard way to overcome this difficulty is to further
impose the light-cone gauge.
For string theory on a curved background, however,
the usual conformal gauge and the light-cone gauge
are not necessary compatible, fixing the light-cone gauge
leads to a modification of the conformal gauge condition \cite{MT,MTT}.

Following the analogy with the Green-Schwarz superstring in flat space,
it seems reasonable to use
a light-cone type gauge to address the problems of integrability
and quantization of superstring theory on $\AdS$. As a first step
in this direction, in the present paper we consider the bosonic
part of the superstring theory on $\AdS$ in such a gauge.

As is known,
the direct quantization of superstrings in the {\it near} plane-wave
background can be performed \cite{Callan,PR} by using exact
solvability of the superstring theory on plane waves \cite{METS}.
In particular, the approach undertaken in \cite{Callan} yields (perturbatively) the
string Hamiltonian as a power series in $1/J$. It appears, however, that there is another
choice of a gauge condition which enables one to determine
the bosonic part of the string Hamiltonian as an exact function of
$J$. This gauge condition is similar to the  uniform
gauge used in \cite{KrTs1} which is related to the static gauge by
a 2d duality transformation \cite{KrTs2}. The uniform gauge uses
the gauge freedom to request that the target space-time would
coincide with the world-sheet time, and that one combination of the global
R-symmetry charges would be homogeneously distributed along the
string. The only difference of our gauge choice from the one used
in \cite{KrTs1} is that the authors of  \cite{KrTs1} distribute the total $\S^5$ angular
momentum $J_1+J_2+J_3$, while we only distribute a single component $J=J_3$ of the $\S^5$
angular momentum. Therefore, in our gauge we study a sector of
string states with one angular momentum fixed, and in the large
$J$ limit we should expect to recover the light-cone plane-wave
Hamiltonian. The uniform gauge we use is in fact a proper
Hamiltonian version of the light-cone gauge of \cite{Callan}. In
our consideration we keep two parameters $J$ and $\l$ finite.
By this reason we can study not only short strings which in the
BMN limit, $J\to\infty$, $\l/J^2$ fixed, represent small
fluctuations around the point-like string carrying large momentum
$J$ along the central circle of $\S^5$, but also strings which
wind around a circle of $\S^5$ and remain long even in the BMN
limit. Rigid long string configuration were studied in
\cite{AFRT}. In the sector of short strings the expansion in $1/J$
of our Hamiltonian reproduces the plane-wave Hamiltonian and the
$1/J$ and $1/J^2$ corrections obtained in \cite{Callan, Swanson}.
We believe that it should be possible to take into account the fermionic degrees of freedom
and obtain a finite $J$, $\l$ Hamiltonian in the
uniform gauge for Green-Schwarz superstring on $\AdS$.

Since both parameters $J$ and $\l$ are finite one could try to
consider the strong coupling limit $\l\to \infty$, $J$ fixed.
It turns out that the Hamiltonian does not have any
good large $\l$ expansion neither in the sector of short strings
nor in the sector of long strings. Nevertheless, one can see the
famous $\sqrt[4]{\l}$ leading behavior of the string energy in the
sector of short strings. The long strings, however, are much
heavier in the strong coupling limit, and their energy scales as
$\sqrt{\l}$.

In our gauge the Hamiltonian is of the Nambu type,\footnote{A
similar but different light-cone Hamiltonian of the square root
type was also obtained for some specific choice of the light-cone
coordinates in the second paper of \cite{Callan}.} and it may seem
difficult to obtain a Lax representation for it. On the other
hand, in the conformal gauge the string model is described as a
reduction of the well-known principle sigma model to the coset
space $\AdS$. In the uniform gauge the world-sheet metric is
non-diagonal and depends non-trivially on the physical fields.
Nevertheless, the Lax representation for the principle model on
arbitrary 2d surface can be easily constructed and further used to
derive the Lax pair for the uniform gauge Hamiltonian. This proves
the (kinematical) integrability of the Hamiltonian. We would like
to emphasize that our method is universal and can be applied to
derive a Lax representation for any gauge-fixed Hamiltonian, in
particular, for the ${\rm AdS}_5$ light-cone Hamiltonian obtained
in \cite{MTT}.

The paper is organized as follows. In section 2 we derive the
physical Hamiltonian in the uniform gauge, and express the
world-sheet metric in terms of physical degrees of freedom. In
section 3 we discuss the (near) BMN and strong coupling limits. In
section 4 we obtain the Lax representation for the physical
Hamiltonian. In section 5 we discuss some general properties of
the monodromy matrix. We show, in particular, how some of the results
obtained in \cite{KMMZ} can be re-derived and generalized within
our approach. In appendices we collect some useful formulae. In particular, in appendix B
we specify our general treatment to the case of classical strings moving in
${\mathbb R}\times {\rm S}^3$.


\section{Gauge-fixed Hamiltonian}
In this section we develop the Hamiltonian formalism for strings
in the uniform gauge. We start with describing suitable
parametrizations of the sphere and the AdS spaces. The five-sphere
can be parametrized by five variables: $y_i,i=1,\ldots, 4$ and by
the angle variable $\phi$. In terms of six real embedding
coordinates $Y_A$, $A=1,\ldots, 6$ obeying the condition $Y_A^2=1$
the parametrization reads \bea \nonumber \Y_1\equiv
Y_1+iY_2=\frac{y_1+iy_2}{1+\frac{y^2}{4}}\, &&,~~~~~
\Y_2\equiv Y_3+iY_4=\frac{y_3+iy_4}{1+\frac{y^2}{4}}\, ,~~~~~\\
\nonumber
&&\hspace{-2.5cm}
\Y_3\equiv Y_5+iY_6=\frac{1-\frac{y^2}{4}}{1+\frac{y^2}{4}} \exp(i\phi) \, .
\eea The metric induced on $\S^5$ from the flat metric of the
embedding space is
$$
dY_AdY_A=\left(\frac{1-\frac{y^2}{4}}{1+\frac{y^2}{4}} \right)^2d\phi^2\,
+\frac{dy_idy_i}{(1+\frac{y^2}{4})^2}\, .
$$
Here and below we use the concise notation $y^2=y_iy_i$.
Analogously to describe the five-dimensional AdS space we introduce four coordinates $z_i$
and  the global AdS time $t$. The embedding coordinates $Z_A$, which obey $\eta_{AB}Z^AZ^B=-1$
with the metric $\eta_{AB}=(-1,1,1,1,1,-1)$, are now parametrized as
\bea
\nonumber
\Z_1\equiv Z_1+iZ_2=-\frac{z_1+iz_2}{1-\frac{z^2}{4}}\, ,&&~~~~~
\Z_2\equiv Z_3+iZ_4=-\frac{z_3+iz_4}{1-\frac{z^2}{4}}\, ,~~~~~\\
\nonumber
&&\hspace{-2.5cm}
\Z_3\equiv Z_0+iZ_5 =\frac{1+\frac{z^2}{4}}{1-\frac{z^2}{4}}
\exp(it)\, .
\eea
For the induced metric one obtains
$$
\eta_{AB}dZ^AdZ^B
=-\left(\frac{1+\frac{z^2}{4}}{1-\frac{z^2}{4}} \right)^2dt^2
+\frac{dz_idz_i}{(1-\frac{z^2}{4})^2} \, .
$$
The same parametrization of the $\AdS$ space was also discussed in
the context of $\AdS$ string quantization in the first paper of
\cite{FT} and in \cite{Callan}.

Since we consider closed strings all the fields $Y_A$ and $Z_A$
are assumed to be periodic functions of the world-sheet coordinate
$0\leq \s\leq 2\pi$. Periodicity implies that the angle variable
$\phi$ has to satisfy the constraint: \bea \label{period}
\phi(2\pi)-\phi(0)=-2\pi  m\, , ~~~~~m\in {\mathbb Z}\, . \eea The
integer number $m$ represents the number of times the string winds
around the circle parametrized by $\phi$.

\medskip

Propagation of bosonic string is described by the sigma model
with $\AdS$ target space. The corresponding Lagrangian density reads
 \bea
\label{L} {\cal L} =-\frac{1}{2}\sqrt{\lambda}~ \gamma^{\a\beta}
\left(-\GA \pa_{\a}t\pa_{\beta}t + \frac{\pa_{\a}z_i\pa_{\beta}z_i
}{(1-\frac{z^2}{4})^2} +\GS \pa_{\a}\phi\pa_{\beta}\phi +
\frac{\pa_{\a}y_i\pa_{\beta}y_i }{(1+\frac{y^2}{4})^2}\right)\,
.\eea Here $\gamma^{\a\beta}\equiv \sqrt{-h}h^{\a\beta} $, where
$h_{\alpha\beta}$ is a world-sheet metric with Minkowski signature. We also introduced two
functions $\GA$ and $\GS$: \bea
\GA=\left(\frac{1+\frac{z^2}{4}}{1-\frac{z^2}{4}} \right)^2\, ,
~~~~~~~~~ \GS=\left(\frac{1-\frac{y^2}{4}}{1+\frac{y^2}{4}}
\right)^2 \, . \eea The string tension $\sqrt{\l}$ is related to
the radius $R$ of $\S^5$ (${\rm AdS}_5$) and the slope $\a'$ of the Regge
trajectory as $\sqrt{\l}=\frac{R^2}{\a'}$.

A consistent quantization procedure would require finding the true
dynamical (physical) variables for the string sigma model. The
most elegant way to achieve this goal is to use the Hamiltonian
formulation. Deriving from eq.(\ref{L}) the canonical momenta for
all the fields we recast the Lagrangian in the phase space form
\bea \nonumber {\cal
L}&=&p_{t}\dot{t}+p_{\phi}\dot{\phi}+p_{z_i}\dot{z_i}+
p_{y_i}\dot{y_i}\\
&-& \nonumber \frac{1}{2\sqrt{\l}~\gamma^{\tau\tau}} \left(
\frac{p_t^2}{ \GA}- \frac{p_{\phi}^2}{ \GS}+\l\GA t'^2
-\l\GS\phi'^2
\phantom{-\frac{y_i'^2}{\Big(1+\frac{y^2}{4}\Big)^2}} \right.
\\
\nonumber &-& \left. \Big(1-\frac{z^2}{4}\Big)^2p_{z_i}^2
-\Big(1+\frac{y^2}{4}\Big)^2p_{y_i}^2 -\frac{\l
z_i'^2}{\Big(1-\frac{z^2}{4}\Big)^2} -\frac{\l
y_i'^2}{\Big(1+\frac{y^2}{4}\Big)^2} \right)
\\\label{LH}
&+&\frac{\gamma^{\tau\s}}{\gamma^{\tau\tau}}\left(p_tt'+p_{\phi}\phi'+p_{z_i}z'_i+
p_{y_i}y'_i\right)\, . \eea Here $p_{t}$, $p_{\phi}$, $p_{z_i}$
and $p_{y_i}$ are the canonical momenta conjugate to $t$, $\phi$,
$z_i$ and $y_i$ respectively. The dot means the derivative with
respect to the world-sheet time $\tau$ while prime denotes the
derivative with respect to $\sigma$. Note also that to derive the
formula we used that the definition of $\gamma^{\a\b}$ implies
$\det\gamma^{-1}=-1$. In what follows we will often use the
shorthand notation for pairings: $p_{z_i}z'_i=p_z z'$, etc.

The uniform gauge we want to fix is of the type considered in
\cite{KrTs1,KrTs2}, and consists in imposing the following two
conditions \bea \label{gauge} t=\tau, ~~~~~~~~~~~~~p_{\phi}=J\, .
\eea Equations of motion for the phase space variables are found
from eq.(\ref{LH}). Upon further substitution of the gauge
conditions (\ref{gauge}) some of these equations turn into the
constraints. Solving the constraints allows one to exclude the
gauge degrees of freedom, and obtain the Hamiltonian for physical
variables. Let us now discuss this procedure in more detail.

First by varying the Lagrangian (\ref{LH}) with respect to
$\gamma^{\tau\s}$ we find an equation to determine $\phi'$ (note that
$t'=0$): \bea \label{con0} \phi'= -\frac{1}{J}(p_{z}z'+p_{y}y') \,
.\eea Integrating this equation over $\sigma$ and recalling
eq.(\ref{period}) we obtain a constraint \bea \label{lm} {\mathcal
V}
=  m\, J \, \, , \eea where \bea {\mathcal V}
=\int_{0}^{2\pi}\frac{{\rm d}\s}{2\pi} (p_{z}z'+p_{y}y')
\, . \eea In the string language this residual constraint is
nothing else but the level-matching condition. We will not try to
solve eq.(\ref{lm}) in classical theory, rather, following the
analogy with the flat case, we will require eq.(\ref{lm}) to be
satisfied by physical states of the theory.

The variable $p_t$ conjugate to the global AdS time $t$ is the
density of the space-time energy of string. On the other hand,
fixing $t=\tau$ allows one to identify $-p_t$ with the Hamiltonian
density ${\mathcal H}$ for physical degrees of freedom: \bea
\label{H} {\rm H}=\int_{0}^{2\pi}\frac{{\rm d}\s}{2\pi}{\mathcal
H}\, \, . \eea To obtain $p_t^2 \equiv {\mathcal H}^2$ one
performs variation of  eq.(\ref{LH}) with respect to $\gamma^{\tau\tau}$
and use eq.(\ref{con0}). In this way we find the square of the
Hamiltonian density ${\mathcal H}^2$ \bea \nonumber {\cal H}^2
&=&\frac{\GA}{\GS}J^2 +\frac{\lambda}{J^2}
\GA\GS(p_{z}z'+p_{y}y')^2 +
\\
\nonumber
&+&\GA\Big(1-\frac{z^2}{4}\Big)^2p_{z}^2
+\lambda\frac{\GA}{\Big(1-\frac{z^2}{4}\Big)^2}z'^2\\
\label{final}
&+&\GA\Big(1+\frac{y^2}{4}\Big)^2p_{y}^2
+\lambda\frac{\GA}{\Big(1+\frac{y^2}{4}\Big)^2}y'^2 \, .
\eea
We therefore see that the physical variables of our theory are the eight coordinates
$y$, $z$ and the corresponding conjugate momenta $p_{y}$, and $p_{z}$. They are subject of the
canonical Poisson brackets
\bea
\nonumber
\{p_{z_i}(\sigma),z_j(\sigma')\}&=&2\pi \delta_{ij}\delta(\sigma-\sigma')\\
\label{poisson} \{p_{y_i}(\sigma),y_j(\sigma')\}&=&2\pi
\delta_{ij}\delta(\sigma-\sigma')\, . \eea It is clear that the
gauge-fixed theory is manifestly invariant under the ${\rm SO}(4)\x
{\rm SO}(4)$ subgroup of the R-symmetry group of the string theory. The
Hamiltonian density is given by the square root and, therefore,
the physical Hamiltonian appears to be of the {\it Nambu type}. It
depends on two parameters, $J$ and $\lambda$. Thus, we have
completely described the hamiltonian structure of the theory, the
equation of motion for any function $\Phi$ of physical variables
is given by \bea \label{evol} \dot{\Phi}=\{{\rm H},\Phi\}\, . \eea

Let us note that we have not yet exploit all the information
contained in eq.(\ref{LH}). In particular, equation for $p_t$
allows one to solve for $\gamma^{\tau\tau}$. Indeed, \bea 0=\frac{\delta
{\cal L}}{\delta
p_t}=1-\frac{1}{\gamma^{\tau\tau}}\frac{p_t}{\sqrt{\l}\GA}\, , \eea {\it
i.e.} \bea \label{g00} \gamma^{\tau\tau}=\frac{p_t}{\sqrt{\l}
\GA}=-\frac{\cal H}{\sqrt{\l}\GA}\, . \eea Note again that to
express the r.h.s. of the last formula via the Hamiltonian density we
picked up the negative root of the equation $p_t^2={\mathcal
H}^2$. Such a prescription is dictated by positivity of the
physical Hamiltonian and by agreement with the plane-wave limit as
will be discussed below. Finally, we can exploit an equation for
$\phi$ \bea \frac{d}{d\tau}\frac{\delta {\cal L}}{\delta
\dot{\phi}} -\frac{\delta {\cal L}}{\delta
\phi}=-\pa_{\sigma}\Big(\frac{\sqrt{\l} \GS}{\gamma^{\tau\tau}}\phi'
+J\frac{\gamma^{\tau\s}}{\gamma^{\tau\tau}}  \Big)=0 \eea to solve for the
metric component $\gamma^{\tau\sigma}$. One finds \bea \label{g01}
\gamma^{\tau\s}=f(\tau)\frac{\cal H}{\sqrt{\l}\GA}
+\frac{\sqrt{\l}}{J^2}\GS(p_{z}z'+p_{y}y')\, , \eea where
$f(\tau)$ is an arbitrary function of $\tau$. The presence of this
function signals a residual symmetry. Indeed, in the Lagrangian
(\ref{lm}) we can shift  the ratio
$\frac{\gamma^{\tau\s}}{\gamma^{\tau\tau}}$ by any function $f(\tau)$. On
the solutions of eq.(\ref{lm}) the Lagrangian remains invariant
under this shift. 
The function $f(\tau)$  
plays the role of the Lagrangian multiplier to the level-matching 
constraint ${\mathcal V}$. Thus, keeping a non-trivial $f(\tau)$
requires the following modification of the Hamiltonian
\bea
{\rm H}\to {\rm H}+f(\tau)({\mathcal V}-mJ)\, .
\eea  
Of course, on solutions of the level-matching constraint this 
Hamiltonian coincides with the old one.
In what follows we pick up $f(\tau)=0$, {\it
i.e.}
 \bea
\label{01}
\gamma^{\tau\s}=\frac{\sqrt{\l}}{J^2}\GS(p_{z}z'+p_{y}y')\, .
\eea
Thus, the world-sheet metric is completely determined in terms of physical fields which is
equivalent to solving the Virasoro constraints.
We see that fixing the uniform gauge invokes a non-trivial gravitational field on the physical
world-sheet. One can easily check that the only constraint
$
{\mathcal V}
$, which we left unsolved, commutes with the Hamiltonian: \bea
\{{\rm H}, {\mathcal V}
\}=0\,
\eea
and is, in fact, a generator of residual symmetry that generates rigid rotations: $\s\to\eps \s$:
\bea
\{
{\mathcal V}
,y(\s)\}=y'(\s)\, , ~~~~~~~~
\{
{\mathcal V}
,p(\s)\}=p'(\s)\, . \eea These equations are of the evolution
type, {\it cf.} (\ref{evol}), where now $\sigma$ plays the role of
the (compact) time variable and the ``Hamiltonian'' is ${\mathcal
V}$ . Thus, ${\rm H}$ and ${\mathcal V}$ generate two commuting
Hamiltonian flows of the dynamical variables corresponding to the times
$\tau$ and $\sigma$.

Since the Hamiltonian commutes with the level-matching constraint
$\V$, the physical string states are divided into sectors labeled
by the winding number $m$ because the eigenvalues of $\V$ are
equal to $m J$, in accord with (\ref{lm}). In what follows we will
loosely refer to strings with zero winding number as short
strings, and to strings with $m\neq 0$ as long strings. Let us
note, however, that since we consider strings moving in $\S^5$, a
string with a nonzero winding number in the $\phi$-direction may
still be of a small size if it is located near the pole of $\S^5$
($y^2\sim 4$).

\section{Near BMN and Strong Coupling Limits}
In this section we use the gauge-fixed Hamiltonian to discuss the
BMN limit, $J\to\infty$ with $\lambda/J^2$ fixed, and the strong
coupling limit, $\lambda\to\infty$ with $J$ fixed. In the uniform
gauge we study a sector of string states with one $\S^5$ angular
momentum fixed, therefore, in the BMN limit we should expect to
recover the light-cone plane-wave Hamiltonian, and the $1/J$ and $1/J^2$
corrections obtained in \cite{Callan, Swanson}.

The authors of \cite{Callan} set up a procedure to construct a
perturbative large-curvature expansion of the string Hamiltonian
on $\AdS$ obtained in the light-cone gauge around the pp-wave
background. In particular, they explicitly obtained the quartic
and (even higher \cite{Swanson}) correction to the pp-wave
Hamiltonian and studied the problem of its diagonalization in
quantum theory. The uniform gauge we chose is, in fact, a
proper
Hamiltonian version of the light-cone gauge used in
\cite{BMN,Callan} valid for finite $J,\ \l$. To make a connection
to the pp-wave limit we rescale the coordinates as \bea
\label{rescale} z\to \frac{1}{\sqrt{J}}z,~~~ y\to
\frac{1}{\sqrt{J}}y,~~~p_{z}\to \sqrt{J} p_{z}, ~~~ p_{y}\to
\sqrt{J} p_{y}. \eea This rescaling is a canonical transformation
because it preserves the canonical Poisson brackets
(\ref{poisson}). To write down the rescaled Hamiltonian density we
introduce an effective BMN coupling $\l'$
$$
\lambda'=\frac{\lambda}{J^2}\, .
$$
To take into account the level-matching condition (\ref{lm}) we
write $(p_{z}z'+p_{y}y')^2$ in the Hamiltonian (\ref{final}) as
$\V^2 + (p_{z}z'+p_{y}y')^2_{\ast}$, where $\V = m J$ is the zero mode,
and the second term with the subscript ${\ast}$ represents the terms
depending on non-zero Fourier modes of $p_{z}z'+p_{y}y'$. Then the
square of the density becomes \bea \nonumber {\cal H}^2
&=&\frac{\GA}{\GS}J^2 + \lambda' m^2 J^2\GA\GS  + \lambda'
\GA\GS(p_{z}z'+p_{y}y')^2_{\ast} +
\\
\nonumber &+&J \GA\left(1-\frac{z^2}{4J}\right)^2p_{z}^2
+J\lambda' \frac{\GA}{\left(1-\frac{z^2}{4J}\right)^2}z'^2\\
\label{rH} &+&J\GA\left(1+\frac{y^2}{4J}\right)^2p_{y}^2 +J\l'
\frac{\GA}{\left(1+\frac{y^2}{4J}\right)^2}y'^2 \, ,\eea where the
rescaled functions $\GA$ and $\GS$ are given by \bea
\GA=\left(\frac{1+\frac{z^2}{4J}}{1-\frac{z^2}{4J}} \right)^2\, ,
~~~~~~~~~ \GS=\left(\frac{1-\frac{y^2}{4J}}{1+\frac{y^2}{4J}}
\right)^2. \eea Now we see that there are two principally
different cases: $(i)$ the case of short strings with the winding
number around the circle parametrized by $\phi$ equal to zero,
$m=0$, and $(ii)$ the case of long strings winding around the
circle with $m\neq 0$. Then the large-curvature expansion around
pp-wave is obtained in the sector of short strings by sending
$J\to\infty$, while keeping the BMN coupling $\l'$ finite. The
leading terms of the large  $J$ expansion are \bea \label{ppexp}
{\mathcal H}=J+{\mathcal H}_{\rm pp}+\ldots \, , \eea where the
second term is a density for the pp-wave Hamiltonian \bea
{\mathcal H}_{\rm
pp}=\frac{1}{2}(p_y^2+p_z^2+y^2+z^2+\l'y'^2+\l'z'^2)\, . \eea
Expanding further one can easily check that the terms of order
$1/J$ and $1/J^2$ precisely agree with those found in
\cite{Callan,Swanson}. Thus, the advantage of our approach is that
it allows us to obtain the physical Hamiltonian for finite $J$
and, therefore, to  study its general properties, without
appealing to perturbation theory. Also a perturbative expansion
becomes easy to handle as it is now encoded in the unique
expression (\ref{rH}).

According to (\ref{ppexp}) in the pp-wave limit ${\cal H}\to J$ and, therefore,
\bea
\gamma^{\tau\tau}\to -\frac{1}{\sqrt{\l'}}\, , ~~~~~~~\gamma^{\tau\s}\to 0\,
\eea
which is essentially the flat metric, as it should be. This also motivates our choice
of the sign in eq.(\ref{g00}).

In the sector of short strings the eigenvalues of the quadratic
Hamiltonian ${\mathcal H}_{\rm pp}$ acting on physical states
satisfying the level-matching constraint (\ref{lm}) are of order
1. This is the reason why the $1/J$ perturbative expansion can be
used in the sector of short strings. On the other hand in the
sector of long strings with the winding number $m\neq 0$ the large
$J$ expansion cannot be used because in that case the
corresponding quadratic Hamiltonian has large eigenvalues of order
$J$ on the physical states, that makes the formal $1/J$ expansion
meaningless. A proper large $J$ expansion in the sector of long
strings requires first to find a classical solution\footnote{A
large class of long string configurations was found in the second
paper of \cite{AFRT}.} satisfying the level-matching condition
(\ref{lm}), and then expand around this solution following the
lines discussed in \cite{FT}.

It is also of interest to consider the strong coupling limit,
$\lambda\to\infty$ with $J$ fixed. In this case we rescale the
coordinates as \bea \label{screscale} z\to
\frac{1}{\sqrt[4]{\l}}z,~~~ y\to
\frac{1}{\sqrt[4]{\l}}y,~~~p_{z}\to \sqrt[4]{\l} p_{z}, ~~~
p_{y}\to \sqrt[4]{\l} p_{y}. \eea This rescaling is clearly a
canonical transformation.\footnote{This rescaling also induces the
rescaling (\ref{rescale}) if $\lambda'$ is finite.} Then the
square of the density takes the form \bea \nonumber {\cal H}^2
&=&\frac{\GA}{\GS}J^2  + \frac{\lambda}{J^2}
\GA\GS(p_{z}z'+p_{y}y')^2 +
\\
\nonumber &+&\sqrt{\l}
\GA\left(1-\frac{z^2}{4\sqrt{\l}}\right)^2p_{z}^2
+ \sqrt{\l}\frac{\GA}{\left(1-\frac{z^2}{4\sqrt{\l}}\right)^2}z'^2\\
\label{strH}
&+&\sqrt{\l}\GA\left(1+\frac{y^2}{4\sqrt{\l}}\right)^2p_{y}^2
+\sqrt{\l} \frac{\GA}{\left(1+\frac{y^2}{4\sqrt{\l}}\right)^2}y'^2
\, ,\eea where the rescaled functions $\GA$ and $\GS$ are given by
\bea
\GA=\left(\frac{1+\frac{z^2}{4\sqrt{\l}}}{1-\frac{z^2}{4\sqrt{\l}}}
\right)^2\, ,~~~~~~~~~
\GS=\left(\frac{1-\frac{y^2}{4\sqrt{\l}}}{1+\frac{y^2}{4\sqrt{\l}}}
\right)^2. \eea In the strong coupling limit the two leading
terms of ${\cal H}^2$ are\footnote{We thank Arkady Tseytlin for an important discussion of this point.} \bea \label{strHH} {\cal H}^2
&\approx&
\frac{\lambda}{J^2}(p_zz'+p_yy')^2
+\sqrt{\l} \left({\mathcal H}_{\rm flat} +\frac{(z^2-y^2)(p_zz'+p_yy')^2}{J^2}\right) \, .\eea
Here ${\mathcal H}_{\rm flat}=p_{z}^2 + z'^2
+p_{y}^2 + y'^2$ is the SO(8) invariant light-cone Hamiltonian density 
for string in flat space.   
It is clear from eq.(\ref{strHH})
that there is no well-defined expansion in $1/\sqrt{\l}$ neither
in the sector of short strings nor in the sector of long strings.
Nevertheless, one can see that at strong coupling the energy of
short strings scales as $\sqrt[4]{\l}$ and the energy of long
strings scales as $\sqrt{\l}$. First of all we notice that if
$m\neq 0$ then the Hamiltonian has the following
expansion (assuming $m > 0$) \bea\label{strHk} {\rm H}\approx
\sqrt{\l} m + \frac{1}{2}\int_0^{2\pi} \frac{{\rm d}\s}{2\pi}\,
\frac{{\mathcal H}_{\rm flat}+\frac{1}{J^2}(z^2-y^2)(p_zz'+p_yy')^2}{\frac{1}{J} (p_zz'+p_yy')}\, . \eea Due to
the non-polynomial structure of the second term the expansion
cannot be used in practice to develop perturbation theory in
$1/\sqrt{\l}$. However, one can see that the contribution of the
second term is subleading, and, therefore, the energy of a generic
long string scales as $\sqrt{\l} m$ in the strong coupling limit.

For $m =0$ even such an expansion as eq.(\ref{strHk}) becomes
impossible. The Hamiltonian takes the form \bea\label{strHo} {\rm
H}\approx \int_0^{2\pi} \frac{{\rm d}\s}{2\pi}\,
\sqrt{\frac{\lambda}{J^2} (p_{z}z'+p_{y}y')^2 +\sqrt{\l}
\left( {\mathcal H}_{\rm flat}+\frac{(z^2-y^2)(p_zz'+p_yy')^2}{J^2}\right)}\, .\eea This time
the first term  under the square root in eq.(\ref{strHo}) is not
the leading one. Indeed, if we drop the second term, take the
root, and integrate over $\s$ we get zero. We conclude, therefore,
that  in the strong coupling limit the energy of a generic short
string scales as $\sqrt[4]{\l}$. Let us mention, however, that
since we consider strings moving in $\S^5$, strings which are
considered to be long in the $\phi$-direction may be of a small
size, and vice verse. Due to this reason there are states in the
sector of long strings whose energies scale as $\sqrt[4]{\l}$, and
there are states in the sector of short strings whose energies
scale as $\sqrt{\l}$ in the strong coupling limit.

\section{The Lax Representation}
In this section we construct the Lax (zero-curvature) representation for the physical
Hamiltonian eqs.(\ref{H}), (\ref{final}), proving therefore that it defines a classical
integrable system.

The space-time we consider is a coset
\begin{center}
$\AdS$=\raisebox{.7ex}{ ${\rm SO}(4,2)\times {\rm SO}(6)\left/\right.$}
\raisebox{-.7ex}{\hspace{-0.3cm}{$~{\rm SO}(5,1)\times {\rm SO}(5)$}}\,
\end{center}
\noindent and, therefore, the string sigma model must be
intimately connected to sigma models on group and coset manifolds.
Since fixing the uniform gauge leads to appearance of the
gravitational field on the world-sheet it is natural to start with
considering the principle sigma model in presence of a
non-trivial two-dimensional metric.
The field variable of the model is a matrix $g$ and the action reads as
\bea {\rm S}=\frac{1}{2}
\int {\rm d}\tau{\rm d}\sigma \gamma^{\a\b}{\rm
Tr}\Big(\pa_{\a}gg^{-1} \pa_{\b}gg^{-1} \Big) \, . \eea In the case of
the flat world-sheet metric integrability of this model is a
well-studied subject \cite{lup,ZM} and
it is based on constructing the zero-curvature (Lax)
representation for the equations of motion (see also \cite{FR}). It is not difficult to
generalize this construction to the case of an arbitrary
world-sheet metric.

To construct the Lax representation for the principle sigma model with
$\gamma^{\a\b}$ arbitrary (but satisfying $\det \gamma = -1$) let
us introduce a current $A_{\a}$ (here $\a$ is $\s$ or $\tau$):
$$
A_{\a}=\pa_{\a}gg^{-1}\,
$$
and its self- and anti-self dual projections \bea
A^{\pm}_{\a}=(P^{\pm})_{\a}^{~\b}A_{\b},~~~~~~~~
(P^{\pm})_{\a}^{~\b}=\delta_{\a}^{~\b}\mp
\gamma_{\a\rho}\epsilon^{\rho\b} \, . \eea Defining the Lax
operator which depends on a spectral parameter $x$ as \bea
\label{Lax}
D_{\a}=\pa_{\a}-\frac{A^{+}_{\a}}{2(1-x)}-\frac{A^{-}_{\a}}{2(1+x)}\equiv
\pa_{\a}- {\mathcal A}_{\a}(x) \eea one can see that equations of
motion \bea \label{eom} \pa_{\a}(\gamma^{\a\b}A_{\b})=0 \eea are
equivalent to the zero curvature condition \bea \label{zc}
[D_{\a},D_{\b}]=0 \, . \eea

It is now easy to generalize this construction to the coset space in hand.
Let us introduce the following matrix $g$
\bea
g=\left(\begin{array}{cc}
g_{a} & 0 \\
0         & g_{s}
\end{array}
\right) \, .
\eea
Here $g_{a}$ and $ g_{s}$ are the following $4\times 4$ matrices ({\it cf}. the second paper of \cite{AFRT})
\bea
g_{a}=\left(\begin{array}{cccc}
0 & \Z_3 & -\Z_2 & \Z_1^* \\
-\Z_3 & 0 & \Z_1 & \Z_2^* \\
\Z_2 & -\Z_1 & 0 & -\Z_3^* \\
-\Z_1^* & -\Z_2^* & \Z_3^* & 0
\end{array}
\right)\, , ~~~~~~~
g_{s}=\left(\begin{array}{cccc}
0 & \Y_1 & -\Y_2 & \Y_3^* \\
-\Y_1 & 0 & \Y_3 & \Y_2^* \\
\Y_2 & -\Y_3 & 0 & \Y_1^* \\
-\Y_3^* & -\Y_2^* & -\Y_1^* & 0
\end{array}
\right) \, . \eea To define these matrices we use the complex
embedding coordinates $\Z_k$, $k=1,2,3$ for the AdS space and
${\Y}_k$ for the sphere. Let us discuss the properties of these
matrices in more detail.

The matrix $g_{a}$ is an element of the group SU$(2,2)$, {\it i.e.} it
obeys \bea g_{a}^{\dagger}Eg_{a}=E, ~~~~~~~~~~~E={\rm
diag}(-1,-1,1,1),\eea provided the following condition is
satisfied \bea \Z_1^*\Z_1+\Z_2^*\Z_2-\Z_3^*\Z_3=-1\, . \eea
In fact $g_{a}$ describes an embedding of an element of the coset
space ${\rm SO}(4,2)/{\rm SO}(5,1)$ into the group ${\rm SU}(2,2)$
which is locally isomorphic to  ${\rm SO}(4,2)$. We use this isomorphism
to work with $4\times 4$ matrices
rather then with $6\times 6$ ones. Quite analogously, $g_{s}$ is
unitary: $g_{s}g_{s}^{\dagger}=1$
given that the embedding fields satisfy  $\Y_k^*\Y_k=1$. This matrix
describes an embedding of an element of the
coset ${\rm SO}(6)/{\rm SO}(5)$ into ${\rm SU}(4)$ the latter being
isomorphic to ${\rm SO}(6)$.

The next step consists in expressing the embedding coordinates  $\Z_k$ and ${\Y}_k$
in terms of physical coordinates and momenta. In particular, $\Y_3$ contains the
unphysical field $\phi$
whose evolution equation is
\bea
\label{mphi}
\dot{\phi}=\frac{\GA}{\mathcal H}\left(\frac{J}{\GS}-\frac{\l}{J^3}\GS(p_z z+p_y y)^2\right)\, .
\eea
Thus, we have a pair of differential equations, (\ref{con0}) and (\ref{mphi}), to determine
$\phi$ via the physical variables.
Integrating (\ref{con0}) we get
\bea
\label{sphi}
\phi(\sigma,\tau)=\phi(0,\tau)-\frac{2\pi}{J}\int_0^{\s}
\frac{{\rm d}\zeta}{2\pi}(p_zz'+p_{y}y')\, .
\eea
Here $\phi(0,\tau)$ can be found (up to time-independent constant) by substituting
eq.(\ref{sphi}) into eq.(\ref{mphi}). Even without solving for $\phi(0,\tau)$ we observe
that the field $\phi(\sigma,\tau)$ possesses a non-trivial monodromy
\bea
\phi(2\pi,\tau)-\phi(0,\tau)=-\frac{2\pi}{J}
{\mathcal V}
\, . \eea As the consequence, the matrix $g_s$, and, therefore,
$g$ have the monodromy which can be written in the form \bea
g_s(2\pi)=Mg_s(0)M\, , \eea where $M$ is a diagonal matrix \bea
M={\rm diag}\Big(e^{\frac{i\pi}{J}{\mathcal V}},
e^{-\frac{i\pi}{J}{\mathcal V}}, e^{-\frac{i\pi}{J}{\mathcal V}},
e^{\frac{i\pi}{J}{\mathcal V}}\Big)\, . \eea

Using the group element $g$ expressed in terms of physical coordinates and momenta we construct
the Lax connection (\ref{Lax}) which is also block-diagonal
\bea
{\mathcal A}_{\a}=
\left(\begin{array}{cc}
{\mathcal A}^a_{\a} & 0 \\
0         & {\mathcal A}^s_{\a}
\end{array}
\right) \, .
\eea

Now we come to the most important point of our construction. One can check that
the dynamical equations (\ref{eom}) are identically satisfied provided that we
use for the world-sheet metric
our solution (\ref{g00}), (\ref{01})  and differentiate $\phi$ according to
eqs.(\ref{con0}) and (\ref{mphi}).
The calculation is straightforward but rather tedious and, therefore, we
refrain from presenting it here. Perhaps, a simplified proof can be found
by using the approach of \cite{Nicolai}.
Thus, we have found the Lax representation for the Hamiltonian (\ref{H}), (\ref{final}).

Let us note that the Lax connection we obtained is non-local
because it explicitly contains the non-local field $\phi$.
Moreover, the sphere component ${\mathcal A}_{\a}^s$ of the
current ${\mathcal A_{\a}}$
is a quasi-periodic function of $\sigma$, \bea {\mathcal
A}_{\a}^s(2\pi)=M{\mathcal A}_{\a}^s(0)M^{-1}\, . \eea These both
pathologies can be cured as follows. Consideration of the
structure of ${\mathcal A_{\a}}$ shows that the current can be
written in the following way \bea {\mathcal
A}_{\a}^s=M(\s)\hat{\mathcal A}_{\a}^sM(\s)^{-1}\, , \eea where
the $\s$-dependent matrix $M$ is given by \bea M(\s)={\rm
diag}\Big(e^{-\frac{i}{2}\phi(\s,\tau)},
e^{\frac{i}{2}\phi(\s,\tau)}, e^{\frac{i}{2}\phi(\s,\tau)},
e^{-\frac{i}{2}\phi(\s,\tau)}\Big)\, \eea and $\hat{{\mathcal
A}}_{\a}^ß$ is local (does not contain $\phi(\s,\tau)$). Since the
zero-curvature representation (\ref{zc}) is invariant under gauge
transformations we can gauge the non-local $\phi$-dependence away.
The local Lax connection arising in this way is given by \bea
\label{local} \L_{\a}^s=M^{-1}(\s){\mathcal
A}_{\a}^sM(\s)-M^{-1}(\s)\pa_{\a}M(\s)= \hat{{\mathcal
A}}_{\a}^s+\frac{i}{2}\pa_{\a}\phi~\Omega\, , \eea where
$\Omega={\rm diag}(1,-1,-1,1)$. Here the derivatives of $\phi$
should be substituted from eqs.(\ref{con0}) and (\ref{mphi}). The
new Lax connection is a periodic function of $\s$ since it is a
local expression in terms of periodic string coordinates. Let us
also note that the original Lax connection (\ref{Lax}) has poles
at $x=\pm 1$ and vanishes at infinity. The gauge transformed
connection (\ref{local}) has the same poles but does not vanish at
infinity. In particular, the Lax component $\L_{\a}^s$ which
will be used in the next section has the following structure \bea
\L_{\s}^s(x)=\frac{~~~ \L^+}{2(1-x)}+\frac{~~~
\L^-}{2(1+x)}-\frac{i}{2J} (p_zz'+p_yy')\Omega \, . \eea

The (left) Lax representation we consider has also a dual
formulation in terms of the right conserved currents
${\mathcal R}^{\a}=-\gamma^{\a\b}g^{-1}\pa_{\b}g$. The relation between these
two formulations is a gauge transformation by the group element
$g$ together with the change $x\to 1/x$ \cite{Maillet} \bea
g^{-1}D_{\a}g=\pa_{\a}-\frac{{\mathcal R}_{\a}^+}{2\Big(1-\frac{1}{x}\Big)}-
\frac{{\mathcal R}_{\a}^-}{2\Big(1+\frac{1}{x}\Big)}\, . \eea

\section{General Properties of Monodromy}
An important object in the theory of integrable systems is the monodromy matrix
${\rm T}(x)$. It is defined as the path-ordered exponential of the Lax component $\L_{\s}(x)$
\bea
{\rm T}(x)=\mathscr{P}\exp\int_0^{2\pi}{\rm d}\s~ \L_{\s}(x) \, .
\eea
The key property of the monodromy matrix is the time conservation of
all its spectral invariants. The trace
${\rm TrT}(x)$, in particular,  generates
an infinite set of integrals of motion\footnote{One can also define
the monodromy matrix by using the quasi-periodic Lax
connection ${\mathcal A}_{\s}$: ${\rm T}_{\rm nl}(x)=
\mathscr{P}\exp\int_0^{2\pi}{\rm d}\s~ {\mathcal A}_{\s}(x)$. However, the spectral invariants
of ${\rm T}_{\rm nl}$ and ${\rm T}$ will coincide only on solutions of constraint (\ref{lm}).
In general the spectral invariants of ${\rm T}_{\rm nl}(x)$ are not conserved.
}. This stems from the fact that the time
evolution of the monodromy is
of the Heisenberg type
\bea
\label{Heis}
\dot{{\rm T}}(x)=[\L_{\tau}(0,\tau),{\rm T}(x)]\, .
\eea
We also recall the Poisson bracket of  ${\rm T}(x)$ with the constraint ${\mathcal V}$:
\bea
\{{\rm T}(x),{\mathcal V}\}=[\L_{\sigma}(0,\tau),{\rm T}(x)]\, .
\eea
Thus, trace of the monodromy 
as well as all its spectral invariants also Poisson
commute with the remaining constraint.
The Jacobi identity $\{{\rm H},\{{\mathcal V},{\rm T}\}\}+{\rm perm.}$
is satisfied by virtue of the Lax representation for $\mathscr{L}$.

To study the analytic properties of the monodromy it is useful to denote the
eigenvalues of ${\rm T}$ as
$\exp(ip_k(x))$, where in our case $k=1,\ldots, 8$. The function $p_k(x)$ is
known as the quasi-momentum (the Floquet function)
and it plays an important role in the (quantum) inverse scattering method \cite{RS}.

As is well known in the theory of integrable PDEs, the {\it local}
integrals of motion are obtained by expanding the eigenvalues of 
${\rm  T}(x)$ around the poles of the Lax connection, which in
our case are at $x=\pm 1$.
It is, therefore, interesting to look at the first non-trivial integral
arising in the expansion around $x=\pm 1$.

Around $x\to 1$ one can use a regular gauge transformation to
bring $\L^+$ to the diagonal form (up to permutations of
eigenvalues). We find the following result \bea \L^+\to
\frac{i}{2\sqrt{\lambda}}{\rm
diag}\Big(\underbrace{\kappa_{+},-\kappa_{+},\kappa_{+},-\kappa_{+}}_{{\rm
AdS}};\underbrace{\kappa_{+},-\kappa_{+},\kappa_{+},-\kappa_{+}}_{\rm
Sphere}\Big) \, . \eea The fact that all the eigenvalues appear to
be proportional to one and the same value $\kappa_+$ is a
consequence of a peculiar form of the matrices $g_a$ and $g_s$
(and the associated currents ${\L}_{\a}$) -- they have a special
property of being skew-symmetric.  After some computation we find
\bea \nonumber \kappa^2_+&=&
\frac{J^2}{\GS}+\frac{\l}{J^2}\GS (p_yy'+p_zz')^2-2\sqrt{\l}~(p_z
z')
+\Big(1+\frac{y^2}{4}\Big)^2 p_y^2+\frac{\l
y'^2}{\Big(1+\frac{y^2}{4}\Big)^2} \, . \eea This expression can
be also rewritten in a more compact form \bea \label{+}
\kappa_+^2=\frac{{\cal H}^2}{\GA}-\Big[
\Big(1-\frac{z^2}{4}\Big)p_{z_k}+
\frac{\sqrt{\l}~z'_k}{\Big(1-\frac{z^2}{4}\Big)}\Big]^2 \, . \eea
Given this result one can directly verify that the integral \bea
\int_0^{2\pi}\frac{{\rm d}\s}{2\pi} \kappa_+ \eea is conserved.
Interestingly enough it does not coincide with the Hamiltonian
${\rm H}$. However, if we reduce the classical string theory to
the one on ${\mathbb R}\times {\rm S}^5$, which amounts to putting
$z_k=0=p_{z_k}$, this integral becomes the string Hamiltonian.

In complete analogy to the previous consideration we determine the
asymptotic behavior of the monodromy around $x= -1$. We find \bea
\L^-\to \frac{i}{2\sqrt{\lambda}}{\rm diag}
\Big(\underbrace{\kappa_{-},-\kappa_{-},\kappa_{-},-\kappa_{-}}_{{\rm
AdS}};
\underbrace{\kappa_{-},-\kappa_{-},\kappa_{-},-\kappa_{-}}_{\rm
Sphere}\Big) \, , \eea where $\kappa_-$ is \bea \label{-}
\kappa_-^2=\frac{{\cal H}^2}{\GA}-\Big[
\Big(1-\frac{z^2}{4}\Big)p_{z_k}-
\frac{\sqrt{\l}~z'_k}{\Big(1-\frac{z^2}{4}\Big)}\Big]^2 \, . \eea
Thus, we observe that two infinite series of local integrals of
motion obtained by expanding the Lax connection around two
different poles are different, they merge under the reduction of
string motion to ${\mathbb R}\times {\rm S}^5$. Since trace of the
monodromy matrix is gauge invariant our results are not specific
to the uniform gauge we use but also hold, {\it e.g.}, in the
conformal gauge. Therefore, in the general case of $\AdS$ the
asymptotic behavior of quasi-momentum around $x=\pm 1$ is related
to the string energy ${\rm H}$ in a very complicated way, quite
opposite to what happens in cases of string theory on ${\mathbb
R}\times {\rm S}^3$ and ${\mathbb R}\times {\rm S}^5$ \cite{KMMZ}.

Finally, it is interesting to consider the reduction to ${\rm
AdS}_5\times {\rm S}^1$, which corresponds to taking all
$y_k=0=p_{y_k}$. Remarkably, in this case the $\kappa^2_{\pm}$
become the perfect squares and we therefore obtain \bea
\kappa_{\pm}=J\mp \frac{\sqrt{\l}}{J}p_zz'\, . \eea Thus, around
$x\to\pm 1$ the quasi-moment behaves as (up to the sign ambiguity)
\bea p(x)=\frac{1}{x\mp 1}\int_0^{2\pi} \frac{{\rm
d}\s}{2\sqrt{\l}}\Big(J\mp \frac{\sqrt{\l}}{J} p_zz'\Big)+\ldots=
\frac{\pi}{x\mp 1}\Big(\frac{J}{\sqrt{\l}}\mp m \Big)+\ldots \, ,
\eea where we made use of the constraint (\ref{lm}). This
asymptotic expansion perfectly agrees with the one obtained for
the string sigma model on  ${\rm AdS}_3\times {\rm S}^1$ by
Kazakov and Zarembo \cite{KMMZ}.

To complete our discussion of the asymptotic properties of the
quasi-momentum, we also exhibit, in the spirit of \cite{KMMZ},
the asymptotic behavior of $p_k(x)$ around $x\to 0$ and $x\to \infty$.
To this end we assume that the classical solutions we consider
carry only the Cartan (abelian) charges of the unbroken
symmetry group ${\rm SO}(4)\times {\rm SO(4)}$: two AdS charges
$S_1$ and $S_2$, and another two charges
$J_1$ and $J_2$, which together with $J$ are the angular momentum
components of string rotating in $\S^5$.
The explicit form of these charges in terms of physical variables
is given in Appendix A. We also assume the fulfillment
of the constraint (\ref{lm}).


In the case $x\to 0$ we find
\bea
{\rm TrT}(x)=8-x^2\frac{8\pi^2}{\lambda}\Big[{\rm H}^2 +S_1^2+S_2^2+J^2+J_1^2+J_2^2\Big]+\ldots ,
\eea
while the individual quasi-momenta exhibit the following asymptotics (up to shifts by integer multiples of $2\pi$)
\medskip
\begin{center}
\begin{tabular}{ll}
\multicolumn{1}{c}{{\small AdS}~~~~~~} &
\multicolumn{1}{c}{{\small Sphere}} \\
$p_1(x)=x\frac{2\pi}{\sqrt{\l}}({\rm H}+S_1-S_2)$,~~~~~~ &
$p_5(x)=x\frac{2\pi }{\sqrt{\l}}(-J+J_1+J_2)$, \\
$p_2(x)=x\frac{2\pi }{\sqrt{\l}}({\rm H}-S_1+S_2)$,~~~~~~ &
$p_6(x)=x\frac{2\pi }{\sqrt{\l}}(J+J_1-J_2)$,\\
$p_3(x)=x\frac{2\pi }{\sqrt{\l}}(-{\rm H}-S_1-S_2)$,~~~~~~ &
$p_7(x)=x\frac{2\pi }{\sqrt{\l}}(J-J_1+J_2)$,\\
$p_4(x)=x\frac{2\pi }{\sqrt{\l}}(-{\rm H}+S_1+S_2)$,~~~~~~ &
$p_8(x)=x\frac{2\pi }{\sqrt{\l}}(-J-J_1-J_2)$\, .
\end{tabular}
\end{center}

\medskip

Analogously, around $x\to\infty$ we obtain
\bea
{\rm TrT}(x)=8-\frac{8\pi^2}{x^2\lambda}\Big[{\rm H}^2 +S_1^2+S_2^2+J^2+J_1^2+J_2^2\Big]+\ldots ,
\eea
as well as
\begin{center}
\begin{tabular}{ll}
\multicolumn{1}{c}{{\small AdS}~~~~~~} &
\multicolumn{1}{c}{{\small Sphere}} \\
$p_1(x)=\frac{2\pi}{x\sqrt{\l}}(-{\rm H}-S_1+S_2)$,~~~~~~ &
$p_5(x)=\frac{2\pi }{x\sqrt{\l}}(J-J_1-J_2)$, \\
$p_2(x)=\frac{2\pi }{x\sqrt{\l}}(-{\rm H}+S_1-S_2)$,~~~~~~ &
$p_6(x)=\frac{2\pi }{x\sqrt{\l}}(-J-J_1+J_2)$,\\
$p_3(x)=\frac{2\pi }{x\sqrt{\l}}({\rm H}+S_1+S_2)$,~~~~~~ &
$p_7(x)=\frac{2\pi }{x\sqrt{\l}}(-J+J_1-J_2)$,\\
$p_4(x)=\frac{2\pi }{x\sqrt{\l}}({\rm H}-S_1-S_2)$,~~~~~~ &
$p_8(x)=\frac{2\pi }{x\sqrt{\l}}(J+J_1+J_2)$\, .
\end{tabular}
\end{center}
The relative sign ambiguity of $p_k(x)$ has been fixed by requiring that in
the absence of winding 
({\it i.e.} when $m=0$ in (\ref{lm})) 
the
following relation must be satisfied $p_k(1/x)=-p_k(x)$. 

In principle, it is now
straightforward to generalize the results obtained in \cite{KMMZ},
and construct the classical Bethe equations for the string theory on
${\rm AdS}_5 \times \S^1$. Due to the
complicated asymptotic behavior of quasi-momentum around $x\to \pm 1$, eqs.(\ref{+})
and (\ref{-}),
the challenge, however, is to derive the
equations for the string theory on $\AdS$.

\section{Conclusions}
In this paper we developed the Hamiltonian formalism for classical
strings on $\AdS$. The Hamiltonian is obtained in the uniform
gauge and depends on two parameters: the $\S^5$ angular momentum
component $J$ and the string tension $\lambda$. In the large $J$
expansion with the effective BMN coupling $\lambda'$ kept fixed
the Hamiltonian reproduces the plane-wave Hamiltonian and higher
corrections previously found in \cite{Callan,Swanson}. We then
exhibited kinematical integrability of the Hamiltonian (for $J$
and $\lambda$ finite) by explicitly constructing the corresponding
Lax representation. In this respect we note that emergence of an
integrable structure is rather intricate because the Hamiltonian
turns out to be of a non-polynomial (Nambu) type. We further
verified that the asymptotic properties of the quasi-momentum (the
generating function of the integrals of motion) perfectly agree
with the ones obtained earlier for some specific cases
\cite{KMMZ}.

\medskip

Let us now formulate several open problems naturally arising in
our approach. As we have seen, for the general $\AdS$ model the
asymptotics of the quasi-momentum around $x\to \pm 1$ is not
related to the global conserved charges in a simple way. This
appears to be an obstacle in formulating the classical string Bethe equations in full
generality. To get more insight into this problem it is desirable
to analyze the higher local conserved charges arising in the
expansion around $x\to \pm 1$. Alternatively, the local integrals
of motion for the string sigma model can be found by means of the
B\"acklund transform \cite{AS}. It is, therefore, interesting to
construct the B\"acklund equations for the sigma model coupled to
2d gravity and analyze the corresponding conservation laws.

\medskip
The knowledge of the continuous string Bethe equations can be
further used to guess the fundamental Bethe equations which would describe
the quantum string, at least in some asymptotic expansions
\cite{AFS}. Another way to approach the quantization problem is to
find first the separated variables for the classical string
Hamiltonian (\ref{H}). To this end one should investigate the
Poisson structure of the Lax connection $\L$ and establish a
relation to the (dynamical) $r$-matrix approach. We expect, however, that the
Poisson structure will not be ultra-local, {\it i.e.} it will
contain the $\delta'(\s-\s')$ term, as it appears already for the
sigma model in the conformal gauge \cite{Maillet}.

\medskip
To maintain the conformal invariance at the quantum level one
needs to include the fermions. We believe that fermionic degrees
of freedom can be naturally incorporated in the Hamiltonian
approach without spoiling the kinematical integrability. In
particular, the zero-curvature representation for the
Green-Schwarz superstring found in \cite{BPR} could be of use
here. The knowledge of the classical separated variables might
help to approach the formidable problem of finding the separated
variables in the quantum case, see \cite{Sklyanin,Smirnov} for
interesting examples. 

\medskip
The uniform gauge is not the only gauge one
can use to fix the gauge freedom of the string theory on $\AdS$.
In particular, choosing the uniform gauge implies non-zero $J$
and, therefore, leads to missing a sector of string states with
$J=0$. Another interesting gauge condition is the ${\rm AdS}_5$
light-cone gauge proposed in \cite{MTT}. An important advantage of
this gauge is that fermions as well as spinless string states can
be readily taken into account. It would be very interesting to use
our method to derive a Lax pair for the ${\rm AdS}_5$ light-cone
Hamiltonian obtained in \cite{MTT}.

\medskip

Finally, it is of interest to clarify the relation between 
the exact Lax pair we constructed here and the perturbative 
$1/J$ Lax pair recently obtained in \cite{Swan2}. 

\section*{Acknowledgments}
We would like to thank Niklas Beisert, Herman Nicolai, Ari Pankiewicz, Nikolai
Reshetikhin, Matthias Staudacher for
interesting discussions and especially Arkady Tseytlin for
helpful comments on the manuscript. The work of  G.~A. was supported in part
by the European Commission RTN programme HPRN-CT-2000-00131 and by
RFBI grant N02-01-00695. The work of S.~F.~was supported in part
by the 2004 Crouse Award.


\appendix

\section{Equations of Motion and Charges}
Equations of motion for physical variables generated by the Hamiltonian (\ref{H}) are
\bea
\nonumber
\dot{z}_k&=&\frac{\GA}{\cal H}\left[ \left(1-\frac{z^2}{4}\right)^2p_{z_k}
+\frac{\l}{J^2}\GS(p_zz'+p_yy')z'_k\right]\, ,\\
\nonumber
\dot{y}_k&=&\frac{\GA}{\cal H}\left[  \left(1+\frac{y^2}{4}\right)^2p_{y_k}
+\frac{\l}{J^2}\GS(p_zz'+p_yy')y'_k\right] \, ,\\
\nonumber
\dot{p}_{z_k}&=&\frac{z_k}{{\mathcal H}\Big(1-\frac{z^2}{4}\Big)}
\left[  -\frac{{\mathcal H}^2}{\Big(1+\frac{z^2}{4}\Big)}
+\frac{1}{2}\Big(1-\frac{z^2}{4}\Big)^2 p_z^2-\frac{\l z'^2}{2\Big(1-\frac{z^2}{4}\Big)^2}
\right]+\\
\nonumber
&+&\pa_{\s}\left[
\frac{\GA}{\mathcal H}\Big( \frac{\l}{J^2}\GS (p_zz'+p_yy')p_{z_k}+\frac{\l }
{\Big(1-\frac{z^2}{4}\Big)^2}z'_k \Big)  \right]\, , \\
\nonumber
\dot{p}_{y_k}&=&-y_k\frac{\GA}{{\mathcal H}\Big(1+\frac{y^2}{4}\Big)\Big(1-\frac{y^2}{4}\Big)}
\left[
\frac{J^2 }{\GS}-\frac{\l}{J^2}\GS
(p_zz'+p_yy')^2
 \phantom{\frac{y_i'^2\det\gamma^{-1}}{\Big(1+\frac{y^2}{4}\Big)^2}}
\right.\\
\nonumber
&+&\left.\frac{1}{2}\Big(1-\frac{y^2}{4}\Big)\Big( \Big(1+\frac{y^2}{4}\Big)^2
p_y^2-\frac{\l y'^2}{\Big(1+\frac{y^2}{4}\Big)^2} \Big)\right]+\\
&+&\pa_{\s}\left[\frac{\GA}{\mathcal H}\Big( \frac{\l}{J^2}\GS (p_zz'+p_yy')
p_{y_k}+\frac{\l }{\Big(1+\frac{y^2}{4}\Big)^2}y'_k \Big)  \right]\, .
\eea
In particular, the first two equations are used to eliminate $\dot{z}_k$ and
$\dot{y}_k$ from the
current ${\mathcal A}_{\a}$ in favor of the corresponding canonical momenta
$p_{z_k}$ and $p_{y_k}$.

The Hamiltonian (\ref{H}) has ${\rm SO}(4)\times {\rm SO}(4)$
symmetry. It is generated by the following charges:
the Cartan generators $S_1$ and $S_2$ for the ${\rm SO}(4)$
symmetry rotating the AdS coordinates are \bea S_1=\int
_0^{2\pi}\frac{{\rm d}\s}{2\pi}
z_{[1}p_{z_{2]}}\ \, ,~~~~~~~ S_2=\int
_0^{2\pi}\frac{{\rm d}\s}{2\pi}
z_{[3}p_{z_{4]}}\, , \eea while the Cartan
generators $J_1$ and $J_2$ for the ${\rm SO}(4)$ group acting on
the coordinates of the sphere have the form \bea J_1=\int
_0^{2\pi}\frac{{\rm d}\s}{2\pi} y_{[1}p_{y_{2]} }
\ \, ,~~~~~~~ J_2=\int _0^{2\pi}\frac{{\rm
d}\s}{2\pi} y_{[3}p_{y_{4]}}\, . \eea
In these formulae $y_{[1}p_{y_{2]}}=y_1p_{y_2}-y_2p_{y_1}$, etc.

\section{Strings on ${\mathbb R}\times \S^3$}
As we have seen the string Hamiltonian and its Lax representation are rather
complicated. To approach a difficult problem of finding
the separated variables for the Hamiltonian (and their subsequent quantization)
one could start from a smaller subsector, {\it e.g.}, from string theory on ${\mathbb R}\times \S^3$.
This model is not conformal at the quantum level but hopefully it remains
integrable. Unraveling its integrable structure might provide
further insight on the general problem.
In this appendix we collect the relevant formulae to analyze strings
on ${\mathbb R}\times \S^3$ in the Hamiltonian setting.

Reduction to
${\mathbb R}\times \S^3$ consists in
taking $z_k=0=p_{z_k}$ for all $k=1,\ldots 4$ and $y_3=y_4=p_{y_3}=p_{y_4}=0$.
Thus, the physical variables are two coordinates,
$y_1$ and $y_2$, and their conjugate momenta $p_1$ and $p_2$.
The square of the Hamiltonian density becomes
\bea \nonumber {\cal H}^2
=\frac{J^2}{\GS} +\frac{\lambda}{J^2}
\GS(p_{y}y')^2
+\Big(1+\frac{y^2}{4}\Big)^2p_{y}^2
+\frac{\lambda y'^2 }{\Big(1+\frac{y^2}{4}\Big)^2}\, .
\eea
The corresponding Lax connection can be written in terms of $2\times 2$ matrices.
For instance, the
$\s$-component reads as
\bea
\L_{\s}=\frac{~~~\L^+}{2(1-x)}+\frac{~~~\L^-}{2(1+x)}-\frac{i}{2J}(p_{y}y')\s_3 \, ,
\eea
where $\s_3$ is the Pauli matrix and
\bea
\nonumber
\L^{\pm}=\left(
\begin{array}{rr}
\L^{\pm}_{11} & \L^{\pm}_{12} \\
- \L^{\pm *}_{12} & -\L^{\pm }_{11}
\end{array}
\right)\, .
\eea
For $\L^{+}$ we have
\bea
\nonumber
\L^{+}_{11}&=&-\frac{i}{\sqrt{\l}}J+\frac{i}{J}\GS (py')
+\frac{i}{\sqrt{\l}}y_{[1}p_{2]}
+\frac{i}{\left(1+\frac{y^2}{4}\right)^2}y_{[1}y'_{2]} \, , \\
\nonumber
\L^{+}_{12}&=&i\frac{y_1+iy_2}{1-\frac{y^2}{4}}\Big[-\frac{J}{\sqrt{\l}}
+\frac{\GS}{J}(py') \Big]\\
\nonumber
&-&\Big[\frac{1}{\sqrt{\l}}
(p_1+ip_2)+\frac{y'_1+iy'_2}{\left(1+\frac{y^2}{4}\right)^2}
\Big]-\frac{(y_1+iy_2)^2}{4}\Big[\frac{1}{\sqrt{\l }}
(p_1-ip_2)
+\frac{y'_1-iy'_2}{\left(1+\frac{y^2}{4}\right)^2}
\Big]
\eea
and for $\L^{-}$
\bea
\nonumber
\L^{-}_{11}&=&\frac{i}{\sqrt{\l}}J+\frac{i}{J}\GS (py')-\frac{i}{\sqrt{\l}}y_{[1}p_{2]}
+\frac{i}{\left(1+\frac{y^2}{4}\right)^2}y_{[1}y'_{2]} \, , \\
\nonumber
\L^{-}_{12}&=&i\frac{y_1+iy_2}{1-\frac{y^2}{4}}\Big[\frac{J}{\sqrt{\l}}
+\frac{\GS}{J}(py') \Big]\\
\nonumber
&+&\Big[\frac{1}{\sqrt{\l}}
(p_1+ip_2)-\frac{y'_1+iy'_2}{\left(1+\frac{y^2}{4}\right)^2}
\Big]+\frac{(y_1+iy_2)^2}{4}\Big[\frac{1}{\sqrt{\l}}
(p_1-ip_2)
-\frac{y'_1-iy'_2}{\left(1+\frac{y^2}{4}\right)^2}
\Big] \, .
\eea
In these formulae $y_{[1}p_{y_{2]}}=y_1p_{y_2}-y_2p_{y_1}$, etc.

Expansion around the plane-wave limit is constructed by rescaling the
coordinates and momenta according to eqs.(\ref{rescale}). It is not difficult
to compute the monodromy perturbatively in $1/\sqrt{J}$. For instance,
for the matrix elements of ${\rm T}$
\bea
\nonumber
{\rm T}=\left(\begin{array}{cc}
{\rm T}_{11} & {\rm T}_{12} \\
{\rm T}_{21} & {\rm T}_{22}
\end{array}
\right) \eea up to the order $1/J$ we find \bea \nonumber {\rm
T}_{11}=e^{-i\pi\o}-\frac{e^{-i\pi \o}}{J}A\, ,~~~~~~~~~~~~~~~~
{\rm T}_{22}=e^{i\pi \o}-\frac{e^{i\pi \o}}{J}A^*
\eea
and
\bea
\nonumber
{\rm T}_{12}=\frac{e^{-i\pi \o}}{\sqrt{J}}
\int_0^{2\pi}{\rm d}\s b(\s)
e^{i\o \s} \, ,~~~~~~~~~
{\rm T}_{21}=-\frac{e^{i\pi\o}}{\sqrt{J}}
\int_0^{2\pi}{\rm d}\s b^*(\s)
e^{-i\pi\o \s} \, .
\eea
Here we use the notation
\bea
\nonumber
A&=&\int_0^{2\pi}{\rm d}\s\int_0^{\s}{\rm d}\s' ~b(\s)b^*(\s')
e^{i\o(\s-\s')} -\int_0^{2\pi}{\rm d}\s c(\s)\, ,
\eea
where the functions $b(\s)$ and $c(\s)$ are
\bea
\nonumber
b(\s)&=&\frac{1}{\sqrt{\l'}}
\left(-\frac{x}{1-x^2}(i(y_1+iy_2)+p_1+ip_2)-\frac{\sqrt{\l'}}{1-x^2}(y'_1+iy'_2)\right)\, , \\
\nonumber
c(\s)&=&\frac{i}{\sqrt{\l'}}\frac{x}{1-x^2}y_{[1}p_{2]}+\frac{i}{1-x^2}y_{[1}y'_{2]}
+\frac{i}{2}\frac{1+x^2}{1-x^2}(py')\, .
\eea
and we have used the concise notation
\bea
\o=\frac{2}{\sqrt{\l'}}\frac{x}{1-x^2}
\eea
In fact, one can consider $\w$ as the new spectral parameter, the map from
the $x$-plane to the $\o$-plane is two-fold, as
\bea
x=\frac{-1\pm \sqrt{1+\l'\o^2}}{\sqrt{\l'}\o}\, .
\eea
As was discussed in section 5 the local integrals of motion are obtained
by expanding the trace of the monodromy matrix around
the singularities of the Lax connection which are at $x=\pm 1$. However,
analyzing the structure of the monodromy
matrix computed perturbatively in $1/J$ one can recognize that the limit
$x\to \pm 1$ is ill-defined.
This clearly shows that two expansions, $x\to \pm 1$ and $J\to\infty$,
are not permutable.
The model, of course, remains to be integrable, the local charges are
reorganized in a different expansion.

From the results above it is easy to see the appearance of the plane-wave physics.
The conventional way to solve the periodic integrable model is to use separation of
variables \cite{Sklyanin}. If the Poisson bracket
$\{{\rm T}_{12}(x),{\rm T}_{12}(x')\}$
vanishes then the matrix element ${\rm T}_{12}(x)$ can be considered
as a new coordinate. Introduce
a variable
\bea
\nonumber
t(\o)=\int_0^{2\pi}{\rm d}\s b(\s)
e^{i\o \s} \, .
\eea
By using the equations of motion generated by the plane-wave Hamiltonian
\bea
\nonumber
\dot{y}=p, ~~~~~\dot{p}=-y+\l' y''\, ,
\eea
one can easily check that indeed $\{t(\o),t(\o')\}=0$.
Further one finds
 \bea
\nonumber
\{t(\o),t^*(\o')\}=
\Big(-\o\o'+\frac{\o^2}{1-x'^2}
 +\frac{\o'^2}{1-x^2} \Big)\frac{e^{2\pi i(\o-\o')}-1}{\o-\o'}\, .\eea
For $\o$ arbitrary this bracket is not canonical. The canonical bracket arises when $\o$ is an
integer. In this case we have
\bea
\nonumber
\{t(\o),t^*(\o')\}=-2\pi i \o^2
\sqrt{1+\o^2\l'}\delta(\o-\o')\, , \eea {\it i.e.} the canonical (separated)
variables are
\bea
\nonumber
a(\o)=\frac{t(\o)}{i\o\sqrt[4]{1+\o^2\l'}}
\eea with the bracket
\bea
\nonumber
\{a^*(\o'),a(\o)\}=-2\pi i
\delta(\o-\o')\, . \eea
It is also interesting to look at the time evolution of $t(\o)\equiv t(x)$.
Expanding (\ref{Heis})
in inverse powers of $\sqrt{J}$ we find at leading order
 \bea
\nonumber
\dot{t}(x)&=&-i\frac{1+x^2}{1-x^2}t(x)-\frac{\sqrt{\l'}}{x}b\Big(\frac{1}{x},0\Big)\Big(e^{2\pi
i \o}-1\Big) \, .\eea
Here we also exhibit the dependence of the function $b(\s)$ on the spectral parameter $x$.
Thus, in the periodic case the dynamics of the
coefficient $t(x)$ is not simple and depends on the
boundary value of fields at some point.
In fact, this is a major obstacle in application of the
inverse scattering method to the periodic case. Roughly speaking,
for rapidly decreasing fields on a line the role of
zero point is played by infinity where the fields vanish and, therefore,
the dynamics of the transition coefficients simplifies.
In the distinguished case of $\o$ integer, however, the unwanted term containing $b(0)$
disappears and the dynamics of $t(x)$ becomes trivial.
In this case the evolution equation can be immediately integrated and we find
the BMN type formula
\bea
t(\o,\tau)=e^{-i\tau\sqrt{1+\l'\o^2}}t(\o,0)\, ,~~~~~~\o\in {\mathbb Z}\, .
\eea
Finally for $\o$ integer we find
\bea {\rm Tr T}=
(e^{i\pi\o}+e^{-i\pi\o})\Big(1-\frac{|t(\o)|^2}{2J}+\ldots\Big)\, . \eea
One can easily see that the conserved quantity $|t(\o)|^2$ is nothing else
but the density of the plane-wave Hamiltonian written in terms of separated variables.
The corresponding quasi-momentum $p(\o)=\arccos(\frac{1}{2}{\rm T})$ has an expansion in powers of 
$1/\sqrt{J}$. Note that for $\w$ non-integer one would get the following $1/J$ expansion
\bea
p(\o)=\pi \o+\frac{e^{-i\pi\o}A+e^{i\pi \o}A^*}{2J|\sin(\pi \o)|}+...\, .
\eea
This expression does not have the limit $\o\to$ integer. Therefore, at integer values of $\o$
the expansion of the quasi-moment changes drastically. Instead of $1/J$ expansion at regular 
(non-integer) values we have $1/\sqrt{J}$ expansion for $\o$ integer-valued.

The perturbative treatment can be pushed to higher orders in $1/\sqrt{J}$.
It is however hardly possible that ${\rm T}_{12}(x)$ would provide the separated variables:
For $J$ finite dynamics of ${\rm T}_{12}(x)$ is complicated and depends on value of the
fields at zero point.

\end{document}